# The Durability and Fragility of Knowledge Infrastructures: Lessons Learned from Astronomy[1]

**Christine L. Borgman**
Department of Information Studies, University of California, Los Angeles, USA
christine.borgman@ucla.edu

**Peter T. Darch**
School of Information Sciences, University of Illinois at Urbana-Champaign, USA
ptdarch@illinois.edu

**Ashley E. Sands**
Department of Information Studies, University of California, Los Angeles, USA
ashleysa@ucla.edu

**Milena S. Golshan**
Department of Information Studies, University of California, Los Angeles, USA
milenagolshan@ucla.edu

**ABSTRACT**
Infrastructures are not inherently durable or fragile, yet all are fragile over the long term. Durability requires care and maintenance of individual components and the links between them. Astronomy is an ideal domain in which to study knowledge infrastructures, due to its long history, transparency, and accumulation of observational data over a period of centuries. Research reported here draws upon a long-term study of scientific data practices to ask questions about the durability and fragility of infrastructures for data in astronomy. Methods include interviews, ethnography, and document analysis. As astronomy has become a digital science, the community has invested in shared instruments, data standards, digital archives, metadata and discovery services, and other relatively durable infrastructure components. Several features of data practices in astronomy contribute to the fragility of that infrastructure. These include different archiving practices between ground- and space-based missions, between sky surveys and investigator-led projects, and between observational and simulated data. Infrastructure components are tightly coupled, based on international agreements. However, the durability of these infrastructures relies on much invisible work – cataloging, metadata, and other labor conducted by information professionals. Continual investments in care and maintenance of the human and technical components of these infrastructures are necessary for sustainability.

[1] This paper is dedicated to A.J. (Jack) Meadows (1934-2016), astronomer, information scientist, and pioneer in scientific communication.

**INTRODUCTION**

Infrastructures, whether for transportation, telecommunications, or scholarly work, are much more fragile than they appear. While some parts have proved durable, such as the Roman aqueducts, most of the roads connecting them have long since crumbled away. Knowledge infrastructures, which are "robust networks of people, artifacts, and institutions" for producing, exchanging, and sustaining knowledge (Edwards, 2010, p. 17), similarly have some durable components such as printed books and the libraries that have stewarded them for centuries. Yet these components are fragile, as funding for research libraries declines and as digital materials fade away (Rumsey, 2016). Infrastructures – transportation and scholarship alike – build on an installed base and are linked with conventions of practice (Star & Ruhleder, 1996). New components emerge to fill gaps, to extend capabilities, or to replace existing infrastructures altogether. Infrastructures develop and evolve, converge or diverge, or fade away when they no longer are needed, funded, or maintained.

For the purposes of this article, "durability" is persistence over time. A particular component tends to be durable when it continues to serve its intended purposes adequately, provided that sufficient resources (financial, material, and human) have been invested in its care and maintenance. "Fragility" similarly describes a component of infrastructure that is subject to failure or degradation, usually due to uncertain availability of the resources necessary to sustain it. A component that has hitherto proved durable nevertheless can become extremely fragile. Infrastructures – and the links between them – require continuous care and maintenance (Borgman, 2000; Star & Strauss, 1999). To regard a component as intrinsically durable would obscure this necessary work.

Knowledge infrastructures for science are evolving in concert with computational and storage capabilities, growth in data production, and scientific policies that require more open access to publications and data. Astronomy is an ideal domain for studying the durability and fragility of knowledge infrastructures, due to its longevity, scale, and transparency. Knowledge of the skies has accumulated over millennia. Today's astronomy databases incorporate star catalogs constructed over the course of centuries. While astronomy is far from a unified field, its boundaries are more apparent than those of most research domains. Astronomers share telescopes, data archives, common software tools, metadata services, and other large investments in infrastructure. Community investments are coordinated internationally, as major telescope missions have partners from multiple countries. The American Astronomical Society provides the unifying function of publishing the primary English-language journals of the field.

This paper explores the durability and fragility of knowledge infrastructures for astronomy, drawing upon a large and long-term study of infrastructures in multiple scientific domains. We frame the problem and provide initial findings, then point to further explications that are under way.

**LITERATURE REVIEW AND BACKGROUND**
A brief survey of knowledge infrastructures in science and astronomy sets the context for this study.

**Knowledge Infrastructures in Science**
All infrastructures are fragile in the long term, as institutions, technologies, social arrangements, and individual stakeholders change over time. Infrastructures may survive due to parts that are durable, to commitments to sustain the capabilities through periods of change, or in some cases due to benign neglect. Particularly useful in thinking about fragility and durability are the eight dimensions of infrastructure identified by Star and Ruhleder (1996, p. 113): Embeddedness, transparency, reach or scope, learned as part of membership, linked with conventions of practice, embodiment of standards, built on an installed base, and becomes visible upon breakdown. Their model originated in a large study of new scientific technologies being introduced to biology. Many scholars subsequently employed these dimensions to study infrastructures (Borgman, 2000, 2015; Bowker, 2005; Edwards et al., 2013).

Infrastructures are necessary to collect, record, and use data, as these activities are embedded in scholarly practice (Bowker, 2005). Similarly, the practices necessary to interpret data and to maintain their scientific value depend on infrastructures (Ribes & Jackson, 2013). Knowledge infrastructures for science are fragile because they have many points of potential failure. Long-term investments in durable parts of infrastructure are necessary, such as in journals, scholarly societies, data archives, and shared technologies. A single point of failure, such as a network router or a central data archive, can disrupt an entire infrastructure. If well designed, however, some infrastructures can be self-healing, whether by re-routing traffic or by finding alternate paths to a solution (Borgman, 2007; Edwards et al., 2013; Van de Sompel, 2013).

Knowledge infrastructures are expensive to construct and maintain because they must support data collection, analysis, use, and access to information over the long term. Infrastructures also exist at multiple levels of scale, which can create tensions between stakeholders with short-, medium-, and long-term goals (Ribes & Finholt, 2009). The value proposition and burden of costs for scientific infrastructures are much debated (Berman & Cerf, 2013; European Commission High Level Expert Group on Scientific Data, 2010).

**Knowledge Infrastructures in Astronomy**
Astronomy, in the most general sense, is the study of the universe beyond the earth's atmosphere. Most astronomers are concerned with celestial objects or physical phenomena; others are concerned with chemical or biological phenomena; and yet others with the history of the universe. These are but a few of the dimensions along which the science varies (Meadows, 1974). While astronomy is often viewed as a quintessential "big science" (Smith, 1992), it also has many "little science" features (Darch & Sands, 2015). Building new instruments, whether on the ground or in space, can take a decade or more from the proposal to "first light." Some kinds of science can be done in a few months or years, if conducted with observing time on extant instruments or with data taken from archives. Infrastructure must adapt continuously, as each new generation of telescopes may produce orders of magnitude more observations than its predecessors (Strauss, 2014).

Telescopes are among the most durable features of the knowledge infrastructures of astronomy, as their scientific usefulness may last for decades. By replacing older cameras and other observing instruments with newer technologies, the scientific life of some telescopes can be extended for many years. Yet even telescopes can be a fragile infrastructure as labor is necessary to ensure that instruments continue to function over time. Funding has to be assembled from multiple public and private sources, each of which can fluctuate over the lifetime of a telescope, instrument, or individual research project. Missions are funded in stages, thus the research and development of a telescope project might be accomplished, but not its construction or subsequent data collection stages. Astronomy relies more heavily on private philanthropy than most scientific endeavors. Until the mid-twentieth century, much of astronomy, at least in the U.S., was privately funded. Astronomers at elite universities had access to large and modern instruments supported by Rockefeller, Carnegie, and other benefactors (McCray, 2004; Williams, 2014). As NASA, the U.S. National Science Foundation, and public agencies in Europe, Japan, China, India, Australia, and elsewhere made greater investments in astronomy, access to telescope time and to data became more equitable (Munns, 2012; McCray, 2004). The number of professional astronomers in positions at universities and other research centers around the world has grown in parallel (DeVorkin & Routly, 1999).

The astronomy community sets its overall priorities for funding through a negotiated process that is published as the decadal survey (Committee on Survey of Surveys: Lessons Learned from the Decadal Survey Process, 2015). This long negotiation process contributes to the durability of astronomy infrastructure by establishing community commitments, at least for decade-long periods. Public funding for astronomy, like most scientific fields, can be a zero-sum game. Commitments made to new missions, telescopes, instruments, data archives, and individual projects are monies not available for other science. When projects ranked highly in one decadal cycle are ranked much lower in the next, their funding may decline precipitously in favor of new endeavors. As new telescopes come online, others may be decommissioned, disrupting the research of those whose science depends on the older instruments. "Big science" projects that depend on new instruments often are in tension with the "little science" projects that can be accomplished with smaller amounts of funding and data (McCray, 2000). Private philanthropy is easier to find for instruments, buildings, or other durable parts of infrastructure than for the essential maintenance of that infrastructure.

The transition from analog to digital astronomy, which occurred from the 1960s through the 1990s, facilitated fundamental changes in scientific practice (McCray, 2004, 2014). Until the advent of modern digital photography, astronomers spent nights on the mountain, exposing glass plates one at a time. With digital capture, astronomers can specify precise timing, exposure, and data rates. A technician on site can confirm settings and adjust calibration to climate conditions, capturing a data stream for the requesting scientists to analyze later. However, some astronomers still prefer to take their own data on the mountain.

Digital capture results in discrete images that can be copied, transferred, and manipulated far more easily than analog data. Astronomy observations, whether taken on glass plates or digital devices, are measurements of the intensity of electromagnetic radiation (e.g., X-rays, visible light) as a function of position on the sky, wavelength, and time. Glass plates, like books, can survive by benign neglect if kept in adequate environmental conditions. Even that is no guarantee of durability – a flood recently threatened the extensive glass plate collection of the Harvard-Smithsonian Center for Astrophysics (Carlisle, 2016). Astronomers, and astronomy librarians, have maintained the durability of observational data by migrating them to new technologies as they appear (Grindlay et al., 2009). As astronomy data collections have grown in size and in number, continual migration has become far more complex and expensive.

**RESEARCH QUESTIONS AND METHODS**

The research reported here explores knowledge infrastructures in astronomy, drawing on our studies of data practices conducted under a series of grants from the National Science Foundation and the Alfred P. Sloan Foundation since 2009. Most of our astronomy research has focused on the Sloan Digital Sky Survey (SDSS) and the Large Synoptic Survey Telescope (LSST), as explained further in Findings. We have asked questions about infrastructure within the SDSS and LSST communities, and also conducted interviews and observations in complementary areas of astronomy. The questions addressed in this paper are these:

- How has astronomy developed, deployed, and managed knowledge infrastructures for their data?
- What factors contribute to the durability and fragility of knowledge infrastructures in astronomy?

Our astronomy work builds upon comparative and longitudinal studies of data practices in scientific and engineering domains, including embedded sensor networks, biology, undersea science, medicine, and physical sciences (Borgman et al., 2015; Borgman, Darch, Sands, Wallis, & Traweek, 2014; Darch et al., 2015; Darch & Sands, 2015; Pasquetto, Sands, Darch, & Borgman, 2016).

We draw on our studies of data practices in astronomy to explore the knowledge infrastructures on which this community depends. Interviews and observations are used to gather information on how astronomers conduct their research, how they generate or acquire data, and how they manage and exploit those data in the short and long term. We also ask specific questions about infrastructure components, relationships among them, and the origin and evolution of those components. Our questions about knowledge infrastructures cut across dimensions such as scale, central or distributed data collection, and characteristics of data management.

| Sites | Interviews | People | Period |
|---|---|---|---|
| SDSS | 136 | 118 | 2009-2016 |
| LSST | 58 | 50 | 2013-2016 |
| Astronomy Infrastructure | 37 | 26 | 2009-2016 |
| Total | 231 | 194 | 2009-2016 |

**Table 1. Data sources used for research reported in this paper**

As shown in Table 1, we draw on interviews, ethnographic participant-observation, and analysis of webpages and other documents. Ethnographic work has been conducted intermittently, from one day to several weeks at a time, over a period of seven years. Interviews are recorded and professionally transcribed. For analytical coding of interview transcripts, field notes, and documents, we used

NVivo 9, a qualitative analysis software package, and analyzed for emergent themes using grounded theory (Glaser & Strauss, 1967).

**FINDINGS**
The findings are organized as follows, and include literature references due to the array of public documents on which we draw. First we present short summaries of the SDSS and LSST projects as a means to explain the role of sky surveys in astronomy knowledge infrastructures. Second, we describe the components of knowledge infrastructures in astronomy that have proved the most durable, with a focus on data management. Third, we identify some of the features of astronomy research that contribute to the fragility of these infrastructures.

**Sky Surveys: Case Studies**
Astronomy sky surveys are research projects to capture uniform data about a region of the sky. They long predate modern telescopes and digital data archives. Early civilizations tracked the night sky throughout the year, creating star catalogs that could be used for purposes such as navigation. Today's knowledge infrastructures in astronomy incorporate historical star catalogs (Genova, 2013).

*Sloan Digital Sky Survey (SDSS)*
The Sloan Digital Sky Survey, named after its largest funder, the Alfred P. Sloan Foundation, is notable for its commitment to timely data releases via a public data archive. SDSS planning began in the 1990s and survey data collection began in 2000, mapping about one-quarter of the night sky with a focus on galaxies, quasars, and stars. A 2.5-meter optical telescope at Apache Point Observatory in New Mexico was designed, built, and deployed for the collection of the SDSS survey data. Multiple instruments on the telescope collected optical and spectroscopic data. The first phase of the SDSS project (SDSS-I) ran from 2000 to 2005; SDSS-II covered 2005 to 2008. Each was funded as an independent project. SDSS-II expanded the scientific goals and broadened the participation. SDSS-III continued with largely new leadership, collaborating institutions, and scientific goals. SDSS-III collected data through summer of 2014, when SDSS-IV began (Ahn et al., 2012; Finkbeiner, 2010; Gray et al., 2005; "Sloan Digital Sky Survey: Home," 2016).

The SDSS data remain heavily used; a July 2016 search of the astronomy section of the SAO/NASA Astrophysics Data System (ADS) yields more than 10,000 papers mentioning "SDSS" in the title or abstract (ADS, 2016). The actual number of papers using SDSS data is probably much higher, given the common practice of reusing data without citing them in publications (Goodman et al., 2014; Pepe, Goodman, Muench, Crosas, & Erdmann, 2014).

SDSS data are held by the investigators for a short proprietary period to process them into a useful scientific form, and then the data are released openly to the world. Individual investigators, small projects, and educators thus have access to high quality data on which to conduct their own research, with or without external funding. The SDSS dataset serves a wider array of users and uses than anticipated by the architects of this influential sky survey. SDSS data have been reused in multiple scientific communities and have become the basis for citizen science projects such as Galaxy Zoo, which led to Zooniverse (Darch, 2011; Zooniverse, 2014). Astronomers tend to view these public engagement efforts as important investments because they help sustain taxpayer support for continued funding.

*Large Synoptic Survey Telescope (LSST)*
The Large Synoptic Survey Telescope (LSST) is a sky survey based on a telescope currently being built in Chile ("LSST project schedule," 2015). LSST is due to launch a decade-long phase of data collection in 2022, generating up to 15 terabytes of data nightly (LSST Science Collaboration et al., 2009). It will provide data for small teams of scientists to answer fundamental questions about multiple topics, including the solar system, near Earth objects, the Milky Way, and the evolution of the universe. LSST is also of significant interest in particle physics, as one scientific goal is to study dark energy (Ivezić et al., 2014).

The National Science Foundation is the primary funder of the LSST, with contributions for particular components from other sources. For instance, the camera is supported by the U.S. Department of Energy, while the telescope's mirrors are funded primarily by private sources. Initial discussions about LSST began in the 1990s, and by 2001 LSST was ranked as the most important ground-based facility in the decadal survey (Committee for a Decadal Survey of Astronomy and Astrophysics; National Research Council, 2010). Research and development began in 2004, and in 2014, NSF approved funding for LSST to transition to the official Construction phase. This transition is accompanied by ramping up the infrastructures for data collection, management, and accessibility. Significant aspects of this data management work are distributed across five sites in the U.S.

The ethos of openness is fundamental to LSST data management principles (Borgman et al., 2014), although subject to negotiation and restrictions. Code used to build LSST data management infrastructure will be open source and globally available; LSST datasets will be openly accessible within the U.S. and Chile. However, external access to the LSST data will be based on agreements negotiated with individual countries.

**Durability of Knowledge Infrastructures in Astronomy**
Astronomy, as an international and distributed scientific endeavor, has made massive investments in knowledge infrastructures over the last several decades. Most obvious are the large telescopes and sky surveys, funded by multiple sources and countries. Here we focus on the less obvious durable components that contribute to data production, management, and stewardship. These include investments in standards, data archives, metadata and discovery systems, and an overall infrastructure fabric. Most of these investments were made in the last few decades, since astronomy became a digital science.

*Data Standards*
Agreements on data standards, developed in the 1970s and widely adopted by the latter 1980s as part of the transition from analog to digital astronomy, underpin many of the later infrastructure developments in astronomy. The Flexible Image Transport System (FITS) is a file format that encodes essential information about the instrument, conditions of observation, wavelength, time, and sky coordinates in a standard data format. FITS adoption enabled astronomers to combine digital records of observations from multiple instruments (Hanisch et al., 2001).

*Data Archives*
Data archives in astronomy are many, varied, and scattered around the world (Committee on NASA Astronomy Science Centers, & National Research Council, 2007). More data appear to originate from space-based than ground-based missions, as discussed further below. Legacy data, such as scans of glass plates, also are becoming more widely available (Grindlay et al., 2009). FITS remains the most common format for observations in these archives. However, our interviewees explain that data standards are a necessary, but far from sufficient, condition for interpreting archived data or for merging data from multiple sources. To use archived data effectively, these scientists require information about the research questions, methods, and observational conditions under which those data were collected. In turn, the people who create and maintain data archives, including the metadata and documentation about them, are essential parts of the knowledge infrastructures for the community. Astronomers often staff help desks, usually on a rotating basis, as both technical and domain knowledge are necessary to exploit data archives for scientific purposes.

*Metadata and Discovery Systems*
Over the last several decades the astronomy community has constructed extensive infrastructures to integrate data archives, publications, and other information artifacts necessary for their science. Three such systems, two in the U.S. and one in France, were initiated between 1970 and 1995.

Databases to catalog celestial objects and other astronomical phenomena mentioned in publications were first established in the early 1970s, building upon historical practice of creating star catalogs. Objects in our galaxy are cataloged in SIMBAD (the Set of Identifications, Measurements, and Bibliography for Astronomical Data), which is based at The Centre de Données Astronomiques de Strasbourg (CDS) in France. Catalogers read new astronomy publications as they appear, creating metadata records for each mentioned celestial object that can be identified (Centre National de la Recherche Scientifique, 2012; Genova, 2013; Perret et al., 2015; "SIMBAD Astronomical Database," 2016). Now that publications are available as digital text, many objects can be identified algorithmically, but some degree of manual cataloging and verification remains essential to the integrity of each of these databases. As of this writing (July 2016), SIMBAD contains identifiers for 8.3 million unique objects that were mentioned in more than 320,000 papers. Objects outside our galaxy are cataloged in the NASA Extragalactic Database (NED), which was founded in the late 1980s (Helou, Madore, Bicay, Schmitz, & Liang, 1990; "NASA/IPAC Extragalactic Database (NED)," 2016). Solar system and planetary data are cataloged in the NASA Planetary Data System (National Aeronautics and Space Administration, Jet Propulsion Laboratory, 2014).

NASA established the Astrophysics Data System (ADS) in 1993 as a means to coordinate access to its many data systems (ADS, 2016). This was a period of rapid technological change, with the World-Wide Web launching about the same time. For some interesting reasons that we will pursue in a later paper, the Smithsonian Astrophysical Observatory / NASA Astrophysics Data System, as it is now known, became a sophisticated bibliographic system, despite its name. ADS contains records of core astronomy publications back to the 19th century, plus has extensive coverage of grey literature (Kurtz et al., 2000, 2005). ADS curates bibliographic records and links between publications, records of celestial objects, and data archives in CDS, NED, and elsewhere (Accomazzi & Dave, 2011; Borgman, 2013; Kurtz et al., 2005). Astronomers use ADS daily to find information, due to its sophisticated searching features, comprehensive coverage, and analytical tools.

Through a series of partnership agreements, ADS, SIMBAD, NED, and CDS are heavily interlinked, offering an array of tools and services for searching, visualizing, and manipulating observational data. Taken together, these four systems establish relatively clear boundaries of astronomy as a science. However, these boundaries are always in flux. For example, LSST expands the boundaries of astronomy by focusing on dark energy through its collaboration with high energy physics. Broader partnerships may disturb the ability of astronomy to maintain these infrastructures – or they may enhance them.

*Infrastructure Fabric*
Astronomers are well aware of their knowledge infrastructures and can describe articulately their strengths, weaknesses, and gaps. In the 2000 decadal survey, the National Virtual Observatory (NVO) rose to the top priority for funding in its category (Astronomy and Astrophysics Survey Committee, 2001; NVO Interim Steering Committee, 2001). The virtual observatory has several names and incarnations. Some refer to the international collaboration known as the International Virtual

Observatory Alliance that coordinates national initiatives (IVOA, 2016). Based at the Space Science Telescope Institute in Baltimore, the NVO developed a series of technologies and standards. The project was intended to provide long-term public funding for building shared infrastructure in astronomy. The U.S. NVO later became the Virtual Astronomical Observatory. In 2014, the assets of the VAO were transferred to NASA (US Virtual Astronomical Observatory, 2014).

Despite the community commitment to the National Virtual Observatory in the decadal survey of 2000, and a scientific board to oversee the initiative, controversy arose quickly. Some welcomed the critical mass of scientific and software expertise at one locale to build shared infrastructure. Others viewed the effort as overly concentrated at one site, and too far removed from the daily practices of scientific end users. The NVO efforts did not rise to a high funding priority in the 2010 decadal survey, nor did other investments in data archives desired by some of our research subjects. Funding for the core development activities of IVOA, NVO/VAO, and other national initiatives has largely disappeared. However, the U.S. VAO legacy and main infrastructure components are currently sustained by the NASA archives at the Infrared Processing and Analysis Center, the High Energy Astrophysics Science Archive Research Center, and the Space Telescope Science Institute (US Virtual Astronomical Observatory, 2014). The VAO still exists as an entity on a voluntary basis, and member institutions continue to participate in the International Virtual Observatory Alliance. However, the virtual observatory has proved to be far less durable than its proponents expected. A much fuller history and analysis of the virtual observatory as knowledge infrastructure is needed, which we defer to a later paper.

**Fragility of Astronomy Knowledge Infrastructure**
While observational data is the common substrate of astronomy, the sustainability of these data varies widely by research specialty, funding sources, uses, and many other factors. Astronomy, like any academic discipline, employs a diverse array of research methods, instruments, technologies, and theories. Research specialties are inherently unstable due to changes in scientific practice, to local differences in naming, and to the ways in which individuals cross boundaries. Large projects draw their teams from disparate specialties, which can exacerbate collaborative frictions. Our explorations to date reveal three dimensions of astronomy research that contribute to the fragility of their knowledge infrastructures. This is not an exhaustive list, but rather a starting point to examine the durability and fragility of astronomy infrastructures.

*Ground vs. Space-Based Missions*
The most fundamental distinction in data practices encountered in our studies of astronomy infrastructure is between projects to build telescopes on the ground and those to launch them into space. Space-based missions, largely funded by NASA in the U.S., invest in long-term data archiving and access as part of the overall project. Examples include the Hubble, Chandra, and James Webb telescopes. Data from space-based missions are archived by NASA science centers for indefinite periods of time, long after the mission itself may have concluded (Committee on NASA Astronomy Science Centers, & National Research Council, 2007). Ground-based missions, largely funded by NSF and private philanthropy in the U.S., may invest substantial project resources in the design and deployment of data archives, but funding for the archive usually ends with funding for the mission. Examples include sky surveys and major instruments such as the observatories at Mauna Kea, Mt. Wilson, and Palomar, to name a few. While funding for astronomy research varies by country and region, the differences between approaches to archiving ground vs. space-based data appear to originate in scientific practice rather than in funding, per se.

Given that astronomy data remain scientifically valuable long after a mission ends, we have asked why data originating in space are privileged over those originating on the ground, especially given that ground-based telescopes long predate space missions. The usual answer is that space-based missions are more expensive (perhaps 50 times more) and thus the cost of curation is a relatively small addition to the total budget. The proportional cost difference does not explain the lack of long-term investment in valuable data from ground-based telescopes, a fact often lamented by astronomers whose work relies on those data sources.

Another factor is that space-based data may be easier to archive, as the instruments need fewer calibration adjustments. The primary calibration occurs before launch. The background sky in space is static, whereas sky and cloud cover on the ground are different each night. Ground-based telescopes require continual cleaning and calibration adjustments in response to observing conditions. New instruments can be added to telescopes on the ground to extend their scientific life. Conversely, most space-based instruments continue to collect data with the hardware resources they had on launch day. However, software can be used to modify calibration, whether on the ground or in space.

Many astronomers mentioned the different organizational cultures of NASA and NSF in response to our questions about investments in data archiving. NASA takes a long view of the data as part of the scientific mission. These are observational data, taken from one instrument, at one time, in one place, and thus cannot be reconstructed. While the same can be said about observational data from ground-based instruments, NSF generally makes smaller and shorter-term grants than does NASA and its international counterparts. With the exception of investments in large-scale facilities such as supercomputers, supercolliders, and similar shared instruments, NSF tends to fund individual projects, and often very small projects (Heidorn, 2008), whereas NASA maintains most of its data within its own data centers. LSST may prove to be an exception among NSF-funded projects, given the emphasis on data, but data collection is still some years in the future.

*Empirical vs. Theoretical Scientific Inquiry*
The distinction between empiricists, also self-referenced as observers, and theorists is far less sharp than that between ground- and space-based missions. Observers use theory or alerts to determine what data to collect, and may generate new theories from empirical investigations. We use the term empiricists to include those who collect their own data from telescopic instruments – some of whom build their own instruments to do so – and those who use data from the archives of ground- or space-based missions. Theorists are astronomers who construct models of phenomena. They may use data from archives to launch their models. Some theorists in our studies claim not to use any data; others consider their simulated data, their input data, the models themselves, or the output of their models to be their data. Simulated data may be structured in the same ways as observational data, enabling them to be analysed by the same sets of tools. Any one individual can be both an observer and a theorist, although most astronomers tend to concentrate in one or the other areas.

Empiricists need to manage the data they use in their own research, which may or may not include data acquired from astronomy databases. Theorists often acquire data from astronomy archives, and they need ways to manage both these data and outputs of their simulations, some of which may be very large. Astronomers who build simulations appear to have infrastructure needs similar to those of modelers in fields such as climate science or turbulence. Models and their outputs are maintained for varying lengths of time, depending on how difficult they are to reproduce.

*Sky Surveys vs. Investigator-Led Inquiry*
The goal of sky surveys is to document specific characteristics of the night sky within a certain range of the electro-magnetic spectrum, using one telescope that may have multiple instruments. While conceived and led by scientific investigators, sky surveys are a different kind of science than investigator-led studies of specific phenomena or celestial objects. The latter may be short or long term, be conducted by one or many investigators, and draw on one or many data sources. Sky surveys and investigator-led inquiries are synergistic. Surveys are systematic efforts to document the night sky. They produce rich sets of observations worthy of "follow up." Later investigator-led projects pursue phenomena identified in the surveys, leading to new findings and theories. Surveys produce far more data and more events of potential scientific interest than that team's survey scientists can pursue themselves. The need for more follow-up investigation about observations in sky surveys is among the arguments often given for open access to astronomy data.

While sky surveys provide a degree of durability for observational data, those same data may become more fragile through reuse by other investigators. Astronomers often acquire data from multiple surveys and other sources to study objects or phenomena. As data are integrated to form new datasets, they support new scientific questions. However, those derived datasets often fail to be sustained. Investigators may hold them as long as they remain useful, and may discard them if too large to maintain. Few astronomy archives can accept derived data that have multiple, poorly documented, or unknown provenance.

**DISCUSSION AND CONCLUSION**
Astronomy has developed, deployed, and managed knowledge infrastructures over long periods of time. Observations collected millennia ago for documenting the movement of stars and planets throughout the year originally served purposes of navigation, religion, and culture. As the science has become more digital, more data-intensive, and more collaborative, those infrastructures, divisions of labor, knowledge, and expertise have evolved. Now those early star catalogs provide continuity in studies of the universe. Modern sky surveys, such as the Sloan Digital Sky Survey and the Large Synoptic Survey Telescope, contribute durability to these knowledge infrastructures by serving as trusted collections that are heavily used by the community.

As astronomy became digital, the community established standards, tools, metadata, and discovery systems to exploit and sustain access to their data. All of these systems and services must be maintained continuously. Metadata and discovery systems such as CDS, NED, and ADS became durable parts of the field's infrastructure over a period of several decades. Particularly notable is the amount of expert human labor necessary to identify and catalog individual celestial objects and types of phenomena. This invisible work creates the links between components of these infrastructures (Borgman, 2000; Star & Strauss, 1999). The process of developing an infrastructure fabric under the rubric of the virtual observatory also reveals the fragility of the larger knowledge infrastructure of the field. Stakeholders endorsed the need for infrastructure investments, but disagreed on matters of centralization, standards, and other features. The durability of CDS, NED, ADS, and other essential components depends on periodic renewal of funding. Directors of these agencies, and the communities who rely on these resources, must explain and argue for the value of these investments on a regular basis.

Despite the durability of the data produced by sky surveys, the astronomy community does not embrace these investments unanimously. Some scientists view large infrastructure investments as monies not spent on smaller projects that produce results on shorter scientific time frames or that employ graduate students and post-doctoral fellows on local projects. Other scientists defend investments in sky surveys and space missions on the basis that they produce observational data that can be mined for a generation. Astronomy is the rare science that has a community mechanism to negotiate these tensions, namely the decadal survey. Consensus is not the same as unanimity, and priorities shift in each decadal review cycle.

The durability and fragility of knowledge infrastructures in astronomy appears partly to be a function of how those infrastructures are deployed in different specialties, and perhaps in different countries and scientific policy regimes. Space-based missions incorporate a long-term commitment to maintaining the value of the observational data collected. Even here, the degree of commitment varies considerably, whether measured by the percentage of project funding devoted to data management, by the number of staff, or by the range of data stewardship services that are provided. Ground-based sky surveys produce archives of observational data that others can use to follow up events and phenomena, and yet do not make the same long-term commitment to maintaining the scientific usefulness of those data beyond the life of the project. From a science policy perspective, the difference in commitment to ground-based and space-based data remains curious. The commitment to maintaining observational data is higher than to simulated data or models, however. Even when observational data are maintained in durable repositories, derivations of those data that result from later reuse may not be sustainable. What is most apparent from our studies is that the degree of human labor devoted to data collection, metadata creation, data curation, data integration, and stewardship is massive and underappreciated (Sands, 2016).

Infrastructure is fragile, even for one of the most durable of sciences – astronomy. The invisible work necessary to maintain individual systems, tools, technologies, standards, and other resources – much of it done by information professionals – may only become visible upon breakdown. Thus the fundamental tenets of infrastructure apply here (Star & Ruhleder, 1996). The durability of the knowledge infrastructure for astronomy is not guaranteed. Constant vigilance remains essential.

## ACKNOWLEDGMENTS

This research was supported by grants from the National Science Foundation (#1145888, C.L. Borgman, PI; S. Traweek, Co-PI, and #0830976), and the Alfred P. Sloan Foundation (#20113194, C.L. Borgman, PI; S. Traweek, Co-PI, and #201514001, C.L. Borgman, PI). We are grateful to the many members of the astronomy community who granted us interviews, access to their laboratories and offices, and provided rare or unpublished documents. We also thank Irene V. Pasquetto and Bernadette M. Randles for commenting on earlier drafts of this paper.